Title

Enhanced brain structure-function tethering in transmodal cortex revealed by high-frequency eigenmodes

Authors

Yaqian Yang[1,3], Zhiming Zheng[2,3,4,5,6,7,8], Longzhao Liu[2,3,4,5,6], Hongwei Zheng[5,9], Yi Zhen[1,3], Yi Zheng[1,3], Xin Wang[2,3,4,5,6,*], Shaoting Tang[2,3,4,5,6,7,8,†]

Affiliations

1. School of Mathematical Sciences, Beihang University, Beijing 100191, China
2. Institute of Artificial Intelligence, Beihang University, Beijing 100191, China
3. Key laboratory of Mathematics, Informatics and Behavioral Semantics (LMIB), Beihang University, Beijing 100191, China
4. State Key Lab of Software Development Environment (NLSDE), Beihang University, Beijing 100191, China
5. Beijing Advanced Innovation Center for Future Blockchain and Privacy Computing，Beihang University, Beijing 100191, China
6. PengCheng Laboratory, Shenzhen 518055, China
7. Institute of Medical Artificial Intelligence, Binzhou Medical University, Yantai 264003, China
8. School of Mathematical Sciences, Dalian University of Technology, Dalian 116024, China
9. Beijing Academy of Blockchain and Edge Computing (BABEC), Beijing 100085, China

Email: * wangxin_1993@buaa.edu.cn
      † tangshaoting@buaa.edu.cn

Abstract

The brain's structural connectome supports signal propagation between neuronal elements, shaping diverse coactivation patterns that can be captured as functional connectivity. While the link between structure and function remains an ongoing challenge, the prevailing hypothesis is that the structure-function relationship may itself be gradually decoupled along a macroscale functional gradient spanning unimodal to transmodal regions. However, this hypothesis is strongly constrained by the underlying models which may neglect requisite signaling mechanisms. Here, we transform the structural connectome into a set of orthogonal eigenmodes governing frequency-specific diffusion patterns and show that regional structure-function relationships vary markedly under different signaling mechanisms. Specifically, low-frequency eigenmodes, which are considered sufficient to capture the essence of the functional network, contribute little to functional connectivity reconstruction in transmodal regions, resulting in structure-function decoupling along the unimodal-transmodal gradient. In contrast, high-frequency eigenmodes, which are usually on the periphery of attention due to their association with noisy and random dynamical patterns, contribute significantly to functional connectivity prediction in transmodal regions, inducing gradually convergent structure-function relationships from unimodal to transmodal regions. Although the information in high-frequency eigenmodes is weak and scattered, it effectively enhances the structure-function correspondence by 35% in unimodal regions and 56% in transmodal regions. Altogether, our findings suggest that the structure-function divergence in transmodal areas may not be an intrinsic property of brain organization, but can be narrowed through multiplexed and regionally specialized signaling mechanisms.

MAIN TEXT

Introduction

The structural connectome shapes and constrains signaling transmission between neuronal populations, giving rise to complex neuronal coactivation patterns that are thought to support perception, cognition, and other mental functions [1]. Understanding the relationship between structure and function is a fundamental question in neuroscience [2]. Numerous models have been proposed to quantify and articulate this relationship, including statistical models, communication models, and biophysical models [3]. A gradually emerging consensus is that functional connections can be inferred by collective, high-order interactions among neural elements. Despite these modeling advances, the structure-function correspondence is relatively moderate, with structural connectivity rarely accounting for more than 50% of the variance in empirical functional connectivity [4].

Recently, studies of regional structure-function relationships, using different approaches, have independently found that the coupling strength systematically varies across the brain [5-7]. Structure and function are tightly coupled in primary sensorimotor areas but diverge in polysensory association areas, gradually decoupling along a macroscale functional gradient from unimodal to transmodal cortex [8, 9]. Such heterogeneous structure-function correspondence raises the possibility that function cannot be completely predicted by structure alone, implying that the observed structure-function decoupling might be a natural consequence of brain hierarchical microscale organization, including cytoarchitecture [5], intracortical myelination [10], and laminar differentiation [11]. One prominent account posits that the rapid evolutionary expansion of the cortical mantle effectively releases association areas from early sensory-motor hierarchies, resulting in great signal variance and weak structure-function relationship in transmodal cortex [12].

Though widely accepted, the corollary that the structure-function relationship may itself be decoupling in transmodal cortex seems to contradict reality. First, it has been widely believed that the brain network is organized under a trade-off between cost minimization and functionality maximization [13]. If structure contributed little to function, it would be unnecessary to invest such substantial material and metabolic resources to white matter construction in transmodal cortex [14-16]. Second, structure-function divergence confers considerable flexibility to structural connectome organization: any wiring diagram that guarantees the connection profile of unimodal cortex would be as good as any other in maintaining normal brain function. However, abundant empirical evidence indicates that structural connections exhibit a high level of consistency and specificity [17-19]. Third, the structure and function of association areas always change simultaneously. The development of human advanced cognitive capabilities was accompanied by pronounced modifications to connections linking association areas [20] and abnormalities in these connections were reported to be associated with many neuropsychiatric disorders [21-23]. Such covariation implies a close correspondence between structure and function in polysensory transmodal areas.

If structure and function are indeed related, why do we observe the decoupling relationship in transmodal cortex? A possible explanation is that current models leave out requisite neurophysiological processes and signaling patterns when linking structural and functional connectivity. Neuromodulation and microstructural variations fundamentally influence how signals are routed, transformed, and integrated on the underlying anatomical backbone, ultimately manifesting as complicated functional connectivity patterns at the macroscale. Specifically, neuromodulation enables the static structural network to support distinct spatiotemporal propagation regimes [24] while local attributes may elicit regional heterogeneity in signaling strategies and transmission events [25]. Nevertheless, most exiting models only embody several putative neurophysiological mechanisms and tend to relate structure and function in the same way across the brain, inherently limiting the extent to which functional connectivity can be explained by structural connections. For instance, geometric and topological relationships (such as Euclidean distance, path length, and communicability) between nodes in the structural network are common and powerful predictors in functional connectivity reconstruction. However, these correlated predictors open the possibility of potentially homogeneous communication patterns, potentially inducing systematic deviation in structure-function coupling strength across the brain.

Hence, it remains a debate whether the structure-function divergence in association cortex is an inherent property of brain organization or a limitation of existing models. Although recent evidence from a machine learning approach demonstrates that structure-function prediction accuracies can be significantly improved [26], it does not provide mechanistic insight into dynamical processes and activation patterns that underlie functional interactions. Whether structure and function are related in transmodal cortex, and if so, what mechanisms and dynamics govern this relationship remain exciting open questions. Here, we aim to shed light on these questions by assessing regional structure-function relationships using distinct frequency-specific spatiotemporal patterns generated by the eigenmode approach [27,28]. First, we demonstrate that low-frequency eigenmodes, which are shown sufficient to capture the essentials of the whole-brain functional network, contribute little to functional connectional profiles of transmodal cortex, leading to systematically decoupling relationships across the brain. Next, we compare the prediction performance of empirical eigenmodes with pseudo-eigenmodes and find that a large number of high-frequency eigenmodes significantly contribute to structure-function mapping. These high-frequency components used to be associated with noisy and random patterns and were overlooked in previous structure-function relationship analyses. However, our findings suggest that they can effectively enhance the explanation of neuronal coactivation patterns in association areas despite the fact that critical information in high-frequency eigenmodes is weak, scattered, and prone to be obscured by excessive background noise. Finally, we show a gradual convergence between structure and function when only using high-frequency eigenmodes as predictors, directly challenging the speculation that the structure-function relationship may itself be decoupling from unimodal to transmodal cortex. These findings indicate that different brain regions utilize specialized parallel signaling protocols, that is, global, persistent signaling patterns preferentially govern the structure-function tethering in unimodal cortex whereas local, transient dynamical processes play dominant roles in functional connectional profiles of transmodal cortex.

## Results
### Laplacian eigenmodes of the structural connectome

To explore how brain function coupled with structure through different signal mechanisms, we applied the eigendecomposition to the structural connectome Laplacian and obtained a set of orthogonal eigenmodes governing frequency-specific spatiotemporal patterns of signal propagation (Fig. 1: left to middle panels). The role of eigenmodes as mediators of information transmission within the brain arises naturally from the network diffusion model, where the eigenvalues closely relate to spatial complexity and persistent time of spreading processes. Specifically, eigenmodes with near-zero eigenvalues, which are referred to as low-frequency eigenmodes, correspond to global and persistent spreading processes whereas eigenmodes with large eigenvalues, which are referred to as high-frequency eigenmodes, correspond to local and transient spreading processes. Moreover, benefiting from their orthogonality, eigenmodes have been used as a parsimonious basis in the prediction of resting-state functional networks. Functional interactions between neuronal populations can be modeled as a weighted linear combination of eigenmodes (Fig. 1: middle to right panels). The coupling strength was quantified as the goodness of fit, i.e., the Pearson correlation coefficient between predicted and empirical functional connectivity. Our methodology followed the eigenmode approach [27-29] with the important difference that we focused on the regional structure-function relationships estimated by the extracted low-frequency and high-frequency components. Details of the analysis were provided in Materials and Methods.

**Regionally heterogeneous roles of Low-frequency eigenmodes**

According to current literatures [7,30], a small number of low-frequency eigenmodes are sufficient to capture the essence of the functional connectivity matrix (See Supplementary Materials, Fig. S1). This observation, however, was made on whole-brain prediction where regional heterogeneity was neglected. It remains unclear what role low-frequency eigenmodes play in regional structure-function relationships.

To address this question, we estimated the regional structure-function relationships using a multilinear regression model that only comprise low-frequency eigenmodes (See Materials and Methods). The resultant coupling strength mirrors the contribution of low-frequency eigenmode to functional connectivity reconstruction. As shown in Fig. 2A, the distribution of coupling strength was broad (from R=0.2 to R=0.8), suggesting highly variable roles of low-frequency eigenmodes in local structure-function prediction. We then examined the spatial distribution of the coupling strengths (Fig. 2B). We found weak structure-function coupling in the bilateral inferior parietal lobule, lateral temporal cortex, precuneus, and inferior and middle frontal gyri. Conversely, strong structure-function coupling resided bilaterally in the visual and primary somatosensory cortices. Aggregating coupling strength by seven resting-state networks proposed by Yeo et al [31], we found structure and function were gradually decoupled from the primary sensorimotor cortex (visual and somatomotor networks) to the transmodal cortex (default mode network), suggesting that the contribution of low-frequency eigenmodes varies systematically across functional systems (Fig. 2C).

To demonstrate that the heterogeneous contribution of low-frequency eigenmodes is not a trivial pattern driven by network size, we calculated the average R of each resting-state network and compared it with the null distribution generated by randomly permuting brain nodes' assignments (10,000 repetitions). As illustrated in Fig. 2D, we found that the coupling strength in visual and somatomotor networks was significantly larger than expected by chance while default mode, limbic, and ventral attention networks exhibited significantly weaker structure-function coupling ($P < 10^{-4}$). We further compared the distribution of well-predicted nodes whose prediction accuracy was higher than the average level with the distribution of seven resting-state network sizes (Fig. 2E). If low-frequency eigenmodes contribute equally to the structure-function coupling across brain regions, these two distributions would exhibit strong similarity. However, we observed an apparent discrepancy between the distributions. We found that 61% of well-predicted nodes were concentrated in visual and somatomotor networks, which greatly exceeded the corresponding network size (37%). In contrast, 27% of well-predicted nodes were observed in default mode, limbic, and ventral attention networks, which was much smaller than the corresponding network size (51%). Finally, we transformed the node-wise R to its z score concerning the null distribution of each resting-state network. Positive values indicate that structure and function are tightly linked by low-frequency eigenmodes whereas negative values indicate that low-frequency eigenmodes play a poor role in predicting functional connectivity. As shown in Fig. 2F, we observed a gradually worsening performance of low-frequency eigenmodes in relating structure to function from unimodal to transmodal cortex.

Collectively, these findings suggest that the contribution of low-frequency eigenmodes to structure-function prediction is not uniform across the brain. The persistent, spatially slow-varying diffusion patterns captured by low-frequency eigenmodes can adequately explain functional connection profiles of unimodal regions. However, they contribute little to the functional connectivity of transmodal regions, leading to the observed structure-function divergence in these regions.

**Structure-function decoupling induced by low-frequency eigenmodes**

Besides the spatially heterogeneous contribution of low-frequency eigenmodes, there exist many other systematic variations [32] in brain organization. Margulies et al [8] reported a cortical organization from unimodal to transmodal cortex, which simultaneously corresponds to a spectrum of increasingly abstract cognitive functions. Here, we associated the regional structure-function relationship estimated by low-frequency eigenmodes with this macroscale functional gradient to examine whether coupling heterogeneities vary along the unimodal-transmodal hierarchy.

We derived the functional gradient (Fig. 3A) following the method in [8] and correlated it with the node-wise coupling strength. As shown in Fig. 3B, we found a negative correlation between coupling strength R and the functional gradient (Pearson ρ=-0.557). To examine whether such anticorrelation is a meaningful feature of the empirical structural connectome, we generated two types of null models. The first one kept the structural connection topology fixed but randomly permuted nodes' positions along the gradient. The second one preserved the original functional gradient but incorporated no structural information except the degree sequence. As shown in Fig. 3C&D, the correlation coefficient between the coupling R and the functional gradient in the empirical data was significantly lower than those generated by the two null models ($P < 10^{-4}$, 10,000 simulations). This observation suggests that structure-function decoupling from unimodal to transmodal cortex is a nontrivial pattern induced by low-frequency eigenmodes. More specifically, if we link brain structure and function only through signaling processes sustained by low-frequency eigenmodes, the resulting structure-function relationships will become increasingly divergent along the unimodal-transmodal hierarchy.

## Critical information in high-frequency eigenmodes

Although whole-brain functional connectivity can be efficiently captured by a few low-frequency eigenmodes, the correspondence between structure and function is far from perfect. It remains a matter of debate whether structure-function divergence in transmodal regions is an inherent property of brain organization or a consequence of neglecting information requisite for precise prediction.

To test the latter possibility, we first examined whether eigenmodes apart from low-frequency components made significant contributions to functional brain connectivity. We applied each eigenmode to whole-brain structure-function prediction and transformed the prediction accuracy R into a z score relative to the null distribution generated by the corresponding pseudo-eigenmode. The pseudo-eigenmodes are the randomization of the original eigenmodes but preserve the spatial smoothness [33]. We found that a large number of eigenmodes with large eigenvalues significantly outperform the corresponding pseudo-eigenmodes in structure-function prediction, although the prediction accuracy decreased steeply from low-frequency to high-frequency domains (See Supplementary Materials, Fig. S2). This observation suggests that high-frequency eigenmodes may contain weak but valuable information for structure-function coupling.

We then investigated how regional functional connectivity emerges from the underlying structure substrate through transient, geometrically complex diffusion patterns captured by high-frequency eigenmodes. For comparison, high-frequency eigenmodes were selected in descending order of their eigenvalues until the prediction accuracy of the selected set equals that of the low-frequency components (See Methods and Materials). In Fig. 4A, we showed that the accuracy R of the regional prediction using models with high-frequency eigenmodes varies from 0.25 to 0.75, suggesting heterogeneous roles of high-frequency eigenmodes across the brain cortex. Furthermore, the spatial distribution of R values indicates a systematic variation in structure-function coupling strength (Fig. 4B). Structure and function are closely aligned in prefrontal and paracentral cortices but decoupled in visual and primary somatosensory cortices, exhibiting a coupling pattern inverse to that induced by the low-frequency eigenmodes. Aggregating prediction R by seven functional systems, we found the default mode network exhibited the strongest structure-function coupling while the visual network exhibited the weakest coupling strength (Fig. 4C).

Similarly, to confirm that the regionally heterogeneous contribution of high-frequency eigenmodes is a nontrivial pattern, we compared the average R of each functional system with the null distribution generated by randomizing network geometry. We found that the coupling strength in ventral attention and default mode networks was significantly higher than expected by chance while the coupling strength in the visual network was significantly lower (Fig. 4D). We further compared the distribution of well-predicted nodes with the distribution of functional network size to rule out the influence of differences in network size. In concert with the previous finding, the proportion of well-predicted nodes in ventral attention and default mode networks is much higher than the corresponding network size (59% > 41%) whereas the reverse is true for the visual network (2% < 15%) (Fig. 4E). Mapping coupling z scores back to individual brain regions, we found that strong structure-function correspondence was concentrated in transmodal regions (Fig. 4F). This observation suggests that the diffusion patterns captured by high-frequency eigenmodes preferentially contribute to the interpretation of functional interactions in transmodal regions.

## Structure-function convergence in transmodal cortex induced by high-frequency eigenmodes

To investigate how local structure-function relationships estimated by high-frequency eigenmodes vary along the unimodal-transmodal hierarchy, we associated the regional coupling strength with the macroscale functional gradient. As shown in Fig. 5A, we found these two measures were positively correlated (Pearson $\rho=0.513$), suggesting that structural and functional connectivity are increasingly coupled from unimodal to transmodal regions under transient and spatially complex signaling protocols. We further compared the empirical correlation coefficient with two null distributions which were respectively generated by randomizing geometry (Fig. 5B) and topology (Fig. 5C). In both cases, the empirical correlation coefficient was significantly larger than the null values ($P < 10^{-4}$, 10,000 simulations). These results deliver novel insights into structure-function tethering, directly challenging the widely held speculation that the structure-function relationship itself may be gradually decoupled along the unimodal-transmodal gradient.

## Enhanced structure-function tethering via introducing high-frequency eigenmodes

As a final step, we sought to shed light on two important questions: Can high-frequency eigenmodes compensate for the critical information neglected in previous structure-function predictions? If so, how is this information distributed in high-frequency eigenmodes?

Firstly, we focused on the structure-function coupling strength estimated by prediction models with and without high-frequency eigenmodes. The brain regions were divided into two groups based on whether their function gradient values were larger than zero, yielding a unimodal group of 590 and a transmodal group of 410. We found that the prediction accuracy R of both unimodal and transmodal groups significantly improved with the addition of high-frequency eigenmodes (t-test, $P < 10^{-4}$), suggesting that high-frequency eigenmodes provide important supplementary information for structure-function tethering (Fig. 6A). The accuracy R increased by 0.25 for the transmodal group, greater than 0.19 for the unimodal group (Fig. 6B). This between-group difference was magnified when absolute increments were converted to percentage increases (56.19%>34.69%). These findings indicate that the information in high-frequency eigenmodes is biased to characterize signaling mechanisms in transmodal regions. Furthermore, we found that the top 10% of brain regions with the highest percentage increases were mostly located in the transmodal cortex, including the inferior parietal cortex, precuneus, insula, cingulate, and lateral prefrontal cortex (Fig. 6C). The coupling strength between structure and function in these regions increased dramatically with the assistance of high-frequency eigenmodes, jumping by 98.6%-196.2%, highlighting the critical role of high-frequency eigenmodes in enhancing the explanation of neuronal coactivation patterns in transmodal areas.

Further, to address the second question, we quantified the percentage increases in R along with the progressive addition of high-frequency eigenmodes. The high-frequency eigenmode added at each step was randomly selected and the adding process was repeated 100 times. The mean and the standard deviation of the percentage increase in R were illustrated in Fig. 6D. We found that the prediction accuracy R in both unimodal and transmodal groups increased steadily as high-frequency eigenmodes were added gradually, suggesting that the information requisite for structure-function prediction is uniformly distributed in high-frequency eigenmodes. It is noteworthy that the growth curve of the transmodal group is steeper than that of the unimodal group, which consolidates the preference of high-frequency eigenmode for interpreting functional interaction in transmodal regions.

**Robustness and reliability**

In previous sections, we have illustrated the regionally heterogeneous roles of different signaling processes sustained by low-frequency and high-frequency eigenmodes in local structure-function prediction. To confirm the robustness and reliability of these results, we replicated the main findings in another four spatial resolutions (68, 114, 219, 448 nodes) and an independent dataset (Human Connectome Project HCP). We observed that spatial distributions of R values were consistent at different spatial resolutions (See Supplementary Materials, Fig. S3). Moreover, the correlation between node-wise coupling strength and the functional gradient was preserved despite the distinct acquisition and processing techniques in different datasets (See Supplementary Materials, Fig. S4). We further adjusted the range of low-frequency and high-frequency eigenmodes to examine the robustness of our findings to the definition of "low-frequency" and "high-frequency". As shown in Supplementary Materials, Fig. S5, the results exhibit high stability across different ranges.

**Discussion**

The imperfect correspondence between structure and function in macroscale brain networks is an ongoing challenge in network neuroscience [3]. The prevailing hypothesis is that structure and function may be gradually untethered along a macroscale functional gradient spanning from unimodal sensory areas to transmodal areas [5,7,11]. In this work, we revisit this hypothesis on the grounds that typically prediction models may neglect signal propagation patterns that are critical for functional interactions in transmodal cortex. To gain a deeper understanding of how functional connectivity emerges from the underlying anatomical substrate, we take into account distinct signaling protocols by decomposing the structure connectome into frequency-specific diffusion patterns captured by orthogonal eigenmodes [34-36]. Concordant with previous findings [5,37], a gradual decoupling between structure and function along unimodal-transmodal hierarchy is reproduced based on low-frequency eigenmodes which are reported as prominent predictors of whole-brain functional connectivity. Next, we show that apart from low-frequency eigenmodes, high-frequency eigenmodes also significantly contribute to structure-function prediction, even though the information they contain is weak and scattered. Unexpectedly, these high-frequency eigenmodes reverse the decoupling pattern between structure and function across the brain, inducing increasingly convergent structure-function relationships along the unimodal-transmodal hierarchy. Finally, we show that with the assistance of high-frequency eigenmodes, the strength of structure-function coupling exhibits dramatic increases in association areas, especially in the inferior parietal cortex, precuneus, insula, cingulate, and lateral prefrontal cortex.

Our work contributes to understanding the link between structural and functional connectivity from a parallel communication perspective. The structure-function relationship has been fruitfully investigated by formulating models of potential communication dynamics, ranging from centralized forms such as the shortest path to decentralized forms such as the random walk [38-40]. However, the correlation of typical predictors such as path length, navigation, and communicability [41] mirrors the homogeneity of potential signal propagation patterns which may drive systematic deviations in structure-function alignment. A key challenge lies in aggregating heterogeneous signaling protocols in a simple and unified framework and articulating their roles in functional interactions among neuronal elements. In our work, a variety of possible signaling processes (orthogonal eigenmodes) were gleaned from the eigendecomposition of the structural Laplacian, with distinct eigenvalues reflecting different frequencies of spatiotemporal patterns of signal spreading. Interregional functional interactions can be interpreted by activating these frequency-specific networked persistent modes in appropriate proportion. This methodology is in line with recent biophysical models which suggest the coexistence of a set of self-sustained, stimulus-selective activity states, with each one storing a memory item for optimal preparation for stimulus processing [42,43]. Studies investigating temporal dynamics of interregional synchrony also suggest that frequency-specific interactions, which form transient frequency-specific networks, modulate cortical computations and information transformation in the brain [44,45].

For the present analysis, we focus on low-frequency and high-frequency eigenmodes that cover two fundamentally different types of signaling patterns, one sustaining persistent and widespread diffusion processes while the other capturing faster and more geometrically complex signal spreading [35,46]. A rich literature supports the notion that brain activity is preferentially expressed in low frequencies [7] and argues that a small number of low-frequency eigenmodes are sufficient to reconstruct the functional network [30,47]. However, this perspective was largely based on the whole-brain prediction where structure-function relationships are assumed to be uniform across the cortex. We show that low-frequency eigenmodes contribute little to structure-function prediction in transmodal areas. In contrast, high-frequency eigenmodes, which are typically associated with noisy and random activation patterns [48,49], could significantly improve the prediction accuracy in these areas. These findings advance the understanding of the roles of different-frequency eigenmodes in structure-function prediction, emphasizing the importance of high-frequency eigenmodes that used to be on the periphery of attention in eigenmode analyses. The significant contribution of low-frequency and high-frequency patterns highlights multiplexed strategies and multiple mechanisms involved in interregional communication [50-52], suggesting that synchrony among neuronal populations results from the aggregation of global, persistent and local, transient signaling patterns. Physiological signals from distributed brain regions compete and cooperate in different frequency bands, manifesting as distinct synchronization patterns to serve flexible cognitive behaviors [53-55]. These various and abundant frequency-specific patterns allow neuronal elements to share and transmit signals through dynamical reconfiguration on multiple timescales, potentially relaxing the restriction on the material and energy cost of the structural connectome [40,56]. Our findings gain valuable insight into how flexible flow of information is achieved, opening the possibility to address the major unsolved question that how static structural connectome supports fast and flexible reconfiguration of functional networks [1].

Furthermore, our findings suggest regionally heterogeneous roles of different signaling mechanisms across the brain. Persistent and global diffusion patterns described by low-frequency eigenmodes predominantly explain functional connectional profiles of primary unimodal regions. Transient and geometrically complex diffusion processes captured by high-frequency eigenmodes instead support functional interactions in association transmodal cortex. These systematic variations in the prediction performance of different diffusion patterns may be induced by latent microstructural configurations and hierarchical organizing principles in the brain, including morphometric similarity, transcription profiles, and laminar differentiation [10,11,25]. The organization of primary areas is strongly constrained by molecular gradients and early activity cascades [57]. Information is step-wise progressively transformed along serial and hierarchical pathways [58]. Such consistent hierarchical property of unimodal cortex may thus elicit widespread and persistent signaling processes that can be captured by low-frequency eigenmodes. In contrast, the rapid expansion of the cerebral cortex detaches polysensory association areas from the canonical sensory-motor hierarchy, resulting in noncanonical circuit organization that lacks consistent laminar projection patterns [12,59]. Such variation in connectivity patterns may alter the way signals are generated, transformed, and integrated, potentially eliciting fundamentally different signaling mechanisms in association areas [60,61]. Association transmodal cortex is configured to bridge widely distributed functional systems and integrate diverse signals from multiple sources [62-64]. Transient and geometrically complex signaling processes captured by high-frequency eigenmodes may enable transmodal cortex to participate in different communication events in a spatially and temporally precise manner, facilitating efficient information routing and flexible state switching in cognitive behaviors. Our findings are also corroborated by the previous work which suggests that primary sensory and motor networks are closely associated with low-frequency connectome harmonics while higher-order cognitive networks match a broader range of frequency spectrum [46].

Low-frequency and high-frequency eigenmodes respectively induce gradually divergent and convergent structure-function relationships along the unimodal-transmodal gradient. These two reverse coupling patterns offer an alternative perspective for understanding the link between structure and function, that is, structural and functional connectivity may be tightly tethered but current models neglect requisite communication dynamics for precise prediction. With the assistance of high-frequency eigenmodes, the tethering between structure and function is enhanced by 35% in unimodal regions and 56% in transmodal regions. In particular, dramatic increases in coupling strength (98.6%-196.2%) appear in the inferior parietal cortex, precuneus, insula, cingulate, and lateral prefrontal cortex, suggesting that structure-function divergence in transmodal areas may not be an inherent property of brain organization. This is in accordance with the recent study [26] which exploits a machine learning approach to achieve a substantially closer structure-function correspondence than previously implied. Although the information in high-frequency eigenmodes is prone to be obscured by background noise and the exact mechanism underlying structure-function association requires further exploration, our findings open a new opportunity to break the glass ceiling on the performance of structure-function prediction. Considering the steady and continuous growth of prediction accuracy with the increasing number of added high-frequency eigenmodes, the information is expected to be uniformly distributed in the high-frequency domain. Meanwhile, the growth curve of transmodal areas is steeper than that of unimodal areas, suggesting that high-frequency patterns have a propensity to explain neuronal coactivation in transmodal cortex. These results provide fundamental references for future work on distilling information from high-frequency eigenmodes to adequately capture the structure-function relationship.

Since high-frequency eigenmodes are usually on the periphery of attention in structure-function prediction due to their association with noisy and random patterns, there may be some concerns about the reliability of the information they provide. We address this problem in the following three directions. First, the averaged node-wise prediction accuracy of high-frequency eigenmodes equals that of low-frequency components (mean R=0.5), roughly concordant with previous structure-function prediction performance [28,29], suggesting that high-frequency eigenmodes can provide sufficient information for an effective structure-function relationship construction. Second, regional structure-function coupling relationships estimated by high-frequency eigenmodes vary in parallel with a macroscale functional gradient [8] which links the cortical organization to an increasingly abstract functional spectrum. This systematic variation demonstrates the functional relevance of the extent to which function couples with structure, suggesting that high-frequency eigenmodes capture meaningful patterns which cannot be ascribed to noise. Finally, our findings exhibit high stability across 4 spatial resolutions (68, 114, 219, 448 nodes), an independently collected dataset (HCP), and different definitions of "low-frequency" and "high-frequency", demonstrating the reliability of critical information contained in high-frequency eigenmodes.

There are some limitations to our work. Although we demonstrate the important roles of high-frequency eigenmodes in structure-function tethering, it is difficult to distill effective information from a large amount of background noise. Meanwhile, the present work is conducted at the group level, however, recent studies demonstrate that structure-function alignments vary with individual differences such as age, sex, and cognitive performance [6,65,66]. Moreover, we represent functional interactions among neuronal elements simply as static and dyadic connectivity networks, neglecting the possibility of temporal dynamics [67] and high-order interactions [68]. Future work could investigate structure-function coupling with more nuanced models enriched with inter-subject variation and high-order interactions.

## Materials and Methods

### Data acquisition

The analyses were performed in two independent datasets. The main dataset was collected by Department of Radiology, University Hospital Center and University of Lausanne (LAU). The dataset [69] was collected from a cohort of 70 healthy participants (27 females, 28.8±9.1 years old). Informed content approved by the Ethics Committee of Clinical Research of the Faculty of Biology and Medicine, University of Lausanne was obtained from all participants. Diffusion spectrum images (DSI) were acquired on a 3-Tesla MRI scanner (Trio, Siemens Medical, Germany) using a 32-channel head-coil. The protocol was comprised of (1) a magnetization-prepared rapid acquisition gradient echo (MPRAGE) sequence sensitive to white/gray matter contrast (1-mm in-plane resolution, 1.2-mm slice thickness), (2) a DSI sequence (128 diffusion-weighted volumes and a single b0 volume, maximum b-value 8,000 s/mm2, 2.2×2.2×3.0 mm voxel size), and (3) a gradient echo EPI sequence sensitive to blood oxygen level-dependent (BOLD) contrast (3.3-mm in-plane resolution and slice thickness with a 0.3-mm gap, TR 1,920 ms, resulting in 280 images per participant). The supplementary analyses were performed in the dataset from the Human Connectome Project (HCP). This dataset consisted of 56 participants. Informed content, including consent to share de-identified data, approved by the Washington University institutional review board was obtained from all participants. All fMRI acquisitions were preprocessed according to HCP-minimal preprocessing guidelines. For more details regarding acquisitions see ref [70,71].

### Structural and functional network construction

Gray matter was divided into 68 brain regions following Desikan–Killiany atlas [72]. These regions were further subdivided into 114, 219, 448, and 1,000 approximately equally sized nodes according to the Lausanne anatomical atlas using the method proposed by [73]. Individual structural networks were constructed using deterministic streamline tractography, initiating 32 streamline propagations per diffusion direction for each white matter voxel [74]. The structural adjacency matrix that preserved the density and the edge-length distribution of the individual participant matrices was then estimated using a group-consensus approach [75-77]. Functional data were pre-processed using routines designed to facilitate subsequent network exploration [78,79]. A group-average functional connectivity matrix was constructed by averaging individual correlation matrices which consist of correlation among regional time series.

### Laplacian eigenmodes

To generate a dissociation of distinct signaling processes, we performed the eigendecomposition of the structural Laplacian. Specifically, we expressed the structural connectome as an undirected, weighted adjacency matrix **A**. Then, the normalized structural Laplacian can be defined as

$$\mathbf{L} = \mathbf{D} - \mathbf{A},$$

where **D** represents the diagonal weighted degree matrix. Following ref. [49], the structural Laplacian was normalized as $\mathbf{L}' = \mathbf{L}/\lambda_{max}$ to preclude the influence of network sizes and densities, where $\lambda_{max}$ indicated the largest eigenvalues of **L**. Through the eigendecomposition of the normalized structural Laplacian $\mathbf{L}'\mathbf{U} = \mathbf{U}\mathbf{\Lambda}$, we obtained a set of orthogonal eigenmodes $\mathbf{u}_k \epsilon \mathbf{U}$ that correspond to distinct spatiotemporal patterns of signal propagation [35,46]. Their eigenvalues $\lambda_k \epsilon \mathbf{\Lambda}$ are closely related to persistent time and spatial complexity signaling processes. Specifically, eigenmodes with near-zero eigenvalues sustain global and persistent diffusion patterns while eigenmodes with large eigenvalues capture geometrically complex spreading processes that delay quickly. Benefiting from their orthogonality, eigenmodes have been used as a parsimonious basis in the prediction of resting-state functional connectivity [34].

### Regional structure-function prediction

The eigenmode approach is considered as a powerful tool for structure-function prediction due to its appealing feature of representing the relationship simply and explicitly [30]. Functional connectivity matrix **F** can be interpreted as the aggregation of activating networked persistent modes captured by eigenmodes in appropriate proportion [28,29], that is,

$$\hat{\mathbf{F}} = \mathbf{U}\mathbf{C}\mathbf{U}^\mathbf{T} = \sum_{k=1}^{n} c_k \mathbf{u}_k \mathbf{u}_k^T,$$

where **C** is a diagonal matrix with elements $c_k$ to be estimated.

The above formulation can also be derived from the network-diffusion model [35]

$$\frac{d\mathbf{x}(t)}{dt} = -\beta \mathbf{L}\mathbf{x}(t),$$

where **x**(t) denotes the time evolution of neural activity and parameter $\beta$ corresponds to the decay rate. It has the analytical solution $\mathbf{x}(t) = e^{-\beta \mathbf{L} t}\mathbf{x}_0$, where $\mathbf{x}_0$ denotes the initial configuration of the diffusion process. Under the hypothesis that the configuration at a critical time $t_{crit}$ evolving from an initial configuration with only region i active is simply the functional connectivity between region i and all other regions [39,30], the whole-brain functional connectivity matrix can be estimated as

$$\hat{\mathbf{F}} = exp(-\beta \mathbf{L} t_{crit}).$$

By eigendecompositing the matrix **L** into $\mathbf{L} = \mathbf{U}\mathbf{\Lambda}\mathbf{U}^\mathbf{T}$, the above equation can be rewritten as $\hat{\mathbf{F}} = \mathbf{U}e^{-\beta \mathbf{L} t_{crit}}\mathbf{U}^\mathbf{T}$. When expressing unknown parameters $e^{-\beta \mathbf{L} t_{crit}}$ as a diagonal matrix **C**, we obtained the same formulation

$$\hat{\mathbf{F}} = \mathbf{U}\mathbf{C}\mathbf{U}^\mathbf{T} = \sum_{k=1}^{n} c_k \mathbf{u}_k \mathbf{u}_k^T.$$

For regional structure-function prediction, the functional connectivity between node i and remaining node j≠i can be expressed as

$$\hat{\mathbf{F}} = \exp(-\beta \mathbf{L} t_{crit})(0|\ldots|e_i|,\ldots|0)$$
$$= \mathbf{U}\mathbf{C}\mathbf{U}^{\mathbf{T}}(0|\ldots|e_i|,\ldots|0)$$
$$= \sum_{k=1}^{n} c_k\, u_{ik} \boldsymbol{u}_k$$
$$= \boldsymbol{b}\boldsymbol{u}_k$$

where $e_i$ denotes the unit vector in the ith direction and $\boldsymbol{b} = (b_1, \ldots, b_N)$ is the vector of parameters that can be estimated using ordinary least squares method (OLS). Local structure-function correspondence is quantified as the goodness of fit, which is computed as the Pearson correlation coefficient R between predicted and empirical functional connectivity.

**Extraction of low-frequency and high-frequency eigenmodes**

It is widely accepted that a small number of low-frequency eigenmodes are sufficient to capture the essence of functional connectivity [47,49]. We, therefore, defined low-frequency eigenmodes as those outperform pseudo-eigenmodes [47] in whole-brain structure-function prediction. Specifically, we constructed pseudo-eigenmodes that preserved the spatial smoothness of empirical eigenmodes following the method in ref. [33]. Next, we sequentially added eigenmodes to structure-function prediction in an order of increasing eigenvalues. At each step, we compared the increase in prediction accuracy R based on empirical eigenmodes with that based on pseudo-eigenmodes. This process was executed until the pseudo-eigenmode outperformed the empirical eigenmode. The resulting eigenmodes were preserved to constitute low-frequency eigenmodes.

As for high-frequency eigenmodes, we started from the eigenmode with the largest eigenvalue, and sequentially added eigenmodes to the prediction model in decreasing order of their eigenvalues until the averaged node-wise prediction accuracy equaled that based on low-frequency components. We selected high-frequency eigenmodes in this way for two considerations. On the one hand, it ensured that high-frequency eigenmodes encompassed sufficient information on structure-function relationships. On the other hand, it benefited a direct comparison between low-frequency and high-frequency eigenmodes. Certainly, the definition of "low-frequency" and "high-frequency" is not absolute and our findings are robust to moderate adjustment of the eigenmode range (See Supplementary Materials, Fig. S5).

**Acknowledgments**

This work is supported by Program of National Natural Science Foundation of China Grant No. 11871004, 11922102, 62141605 and National Key Research and Development Program of China Grant No. 2018AAA0101100, 2021YFB2700304.

**Funding:**
National Natural Science Foundation of China grant 11922102
National Natural Science Foundation of China grant 11871004
National Natural Science Foundation of China grant 62141605
National Key Research and Development Program of China grant 2018AAA0101100
National Key Research and Development Program of China grant 2021YFB2700304

**Author contributions:**
    Conceptualization: Y.Y., X.W.
    Methodology: Y.Y., Z.Z., L.L., X.W.
    Investigation: Y.Y., Z.Z., L.L., H.Z., Y.Z., Y.Z., X.W., S.T.
    Visualization: Y.Y., Y.Z.

Supervision: X.W., S.T.
Writing—original draft: Y.Y.
Writing—review & editing: L.L., X.W., S.T.

**Competing interests:** Authors declare that they have no competing interests.

**Data and materials availability:** The Lausanne dataset is available at https://zenodo.org/record/2872624#.YTR9lI4zaUl [94]. The Human Connectome project dataset is available at https://www.humanconnectome.org/study/hcp-young-adult. Code for the reported analyses is available at https://github.com/yangyaqian007/Enhanced-structure-function-tethering.

**Figures and Tables**

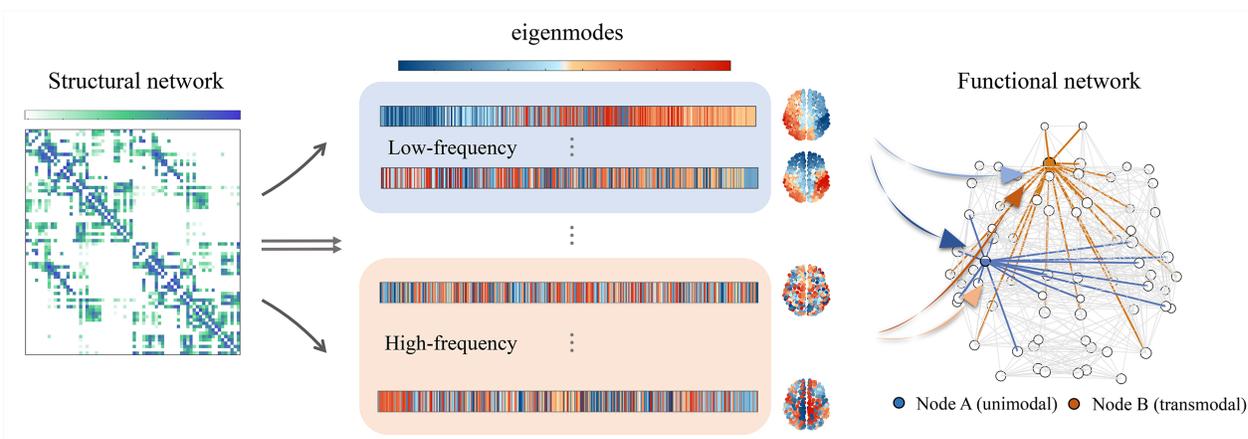

**Fig. 1. Method pipeline.** Through the Laplacian eigendecomposition of the structural network, we obtained a series of orthogonal eigenmodes governing frequency-specific spatiotemporal patterns of signal propagation. The low-frequency and high-frequency components were respectively extracted to predict functional connection profiles of individual brain regions. The structure-function coupling strength was measured as the Pearson correlation coefficient R between predicted and empirical functional connections.

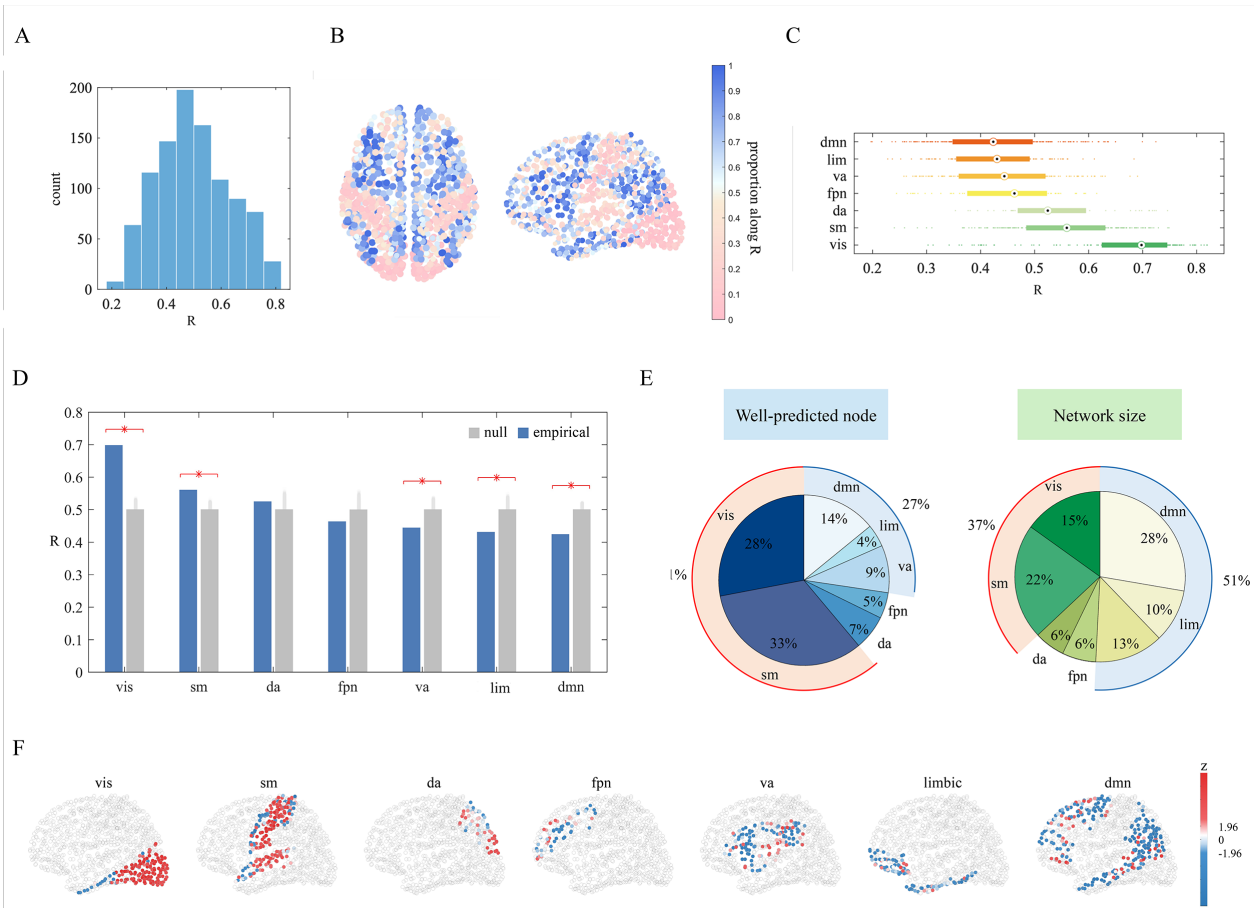

**Fig. 2. Heterogeneous contribution of low-frequency eigenmodes in regional structure-function prediction.** Low-frequency eigenmodes, which are considered to be sufficient to capture the essence of the whole-brain functional network, are exploited to predict functional connection profiles of individual nodes. (**A**), The histogram of node-wise prediction accuracy R. (**B**), The spatial distribution of R. (**C**), Nodes are aggregated by seven resting-state networks (RSNs): visual(vis), somatomotor(sm), dorsal attention(da), frontoparietal(fpn), ventral attention(va), limbic(lim), default mode(dmn) networks. These RSNs are ordered according to their average R-values and the boxplot shows the medians(circles), interquartile ranges(boxes), and outliers(whiskers). (**D**), The empirical average R of each RSN was compared with the null distribution generated by randomly permuting nodes' assignments (10,000 repetitions). Red stars indicate statistically significant differences between empirical and null values. (**E**), The distribution of well-predicted nodes and the distribution of network size across seven RSNs. (**F**), Node-wise R values were transformed into z-scores distributed within each RSN.

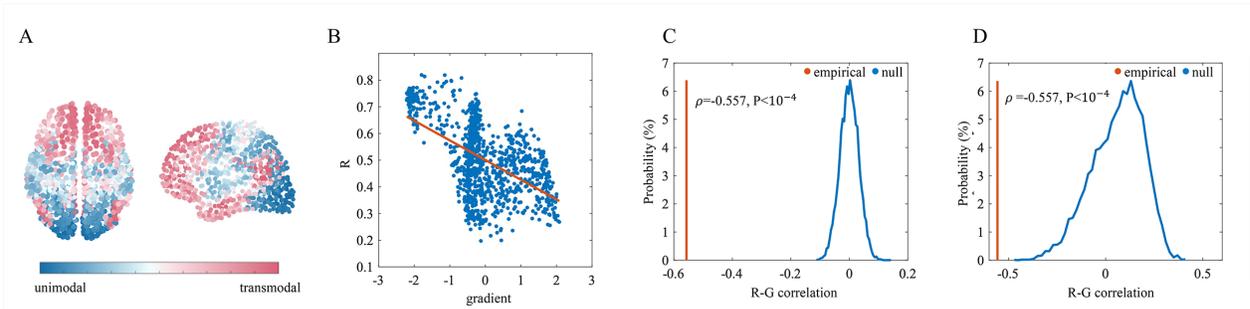

**Fig. 3. Structure-function decoupling along the unimodal-transmodal hierarchy.** (**A**), A macroscale functional gradient spanning from unimodal to transmodal cortex. (**B**), The structure-function coupling strength R is negatively correlated with the macroscale functional gradient. (**C**), Red line: correlation coefficient between R and the functional gradient. Blue curve: the null distribution of correlation coefficients generated by randomly permuting nodes' locations along the functional gradient. (**D**), Red line: correlation coefficient of empirical data. Blue curve: the null distribution generated by randomizing structural architecture.

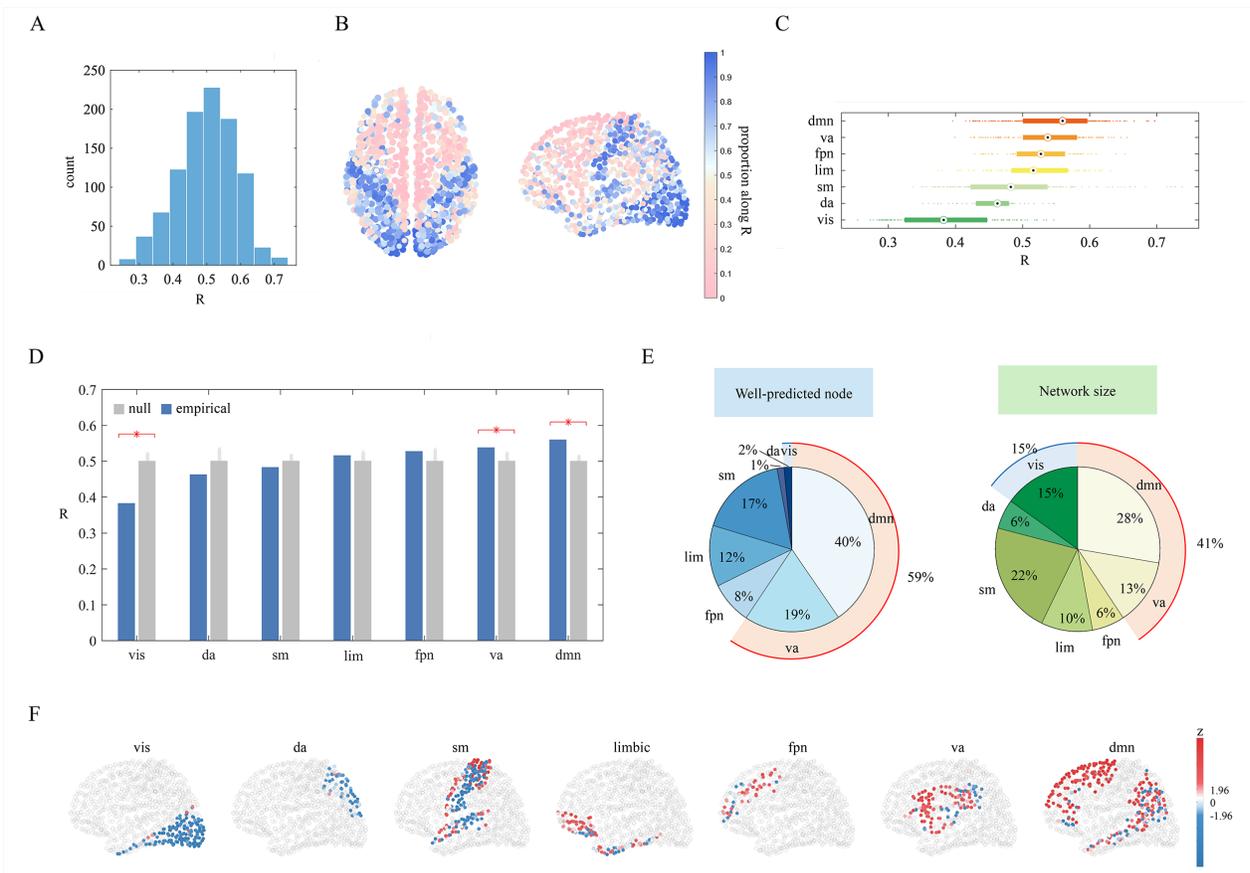

**Fig. 4. The regional prediction contribution of high-frequency eigenmodes.** (**A**), A wide distribution of node-wise prediction accuracy R across 1,000 nodes. (**B**), The corresponding brain map where nodes are colored from blue to red in increasing order of R values. (**C**), Boxplot of structure-function R where RSNs are ordered by the average of R values. (**D**), The average R-value of each RSN is compared with those calculated after random permutation of nodes' assignments (10,000 repetitions); red stars indicate significant differences. (**E**), The distribution of well-predicted nodes across seven RSNs is compared with the distribution of network size. (**F**), The spatial distribution of z scores within each RSN.

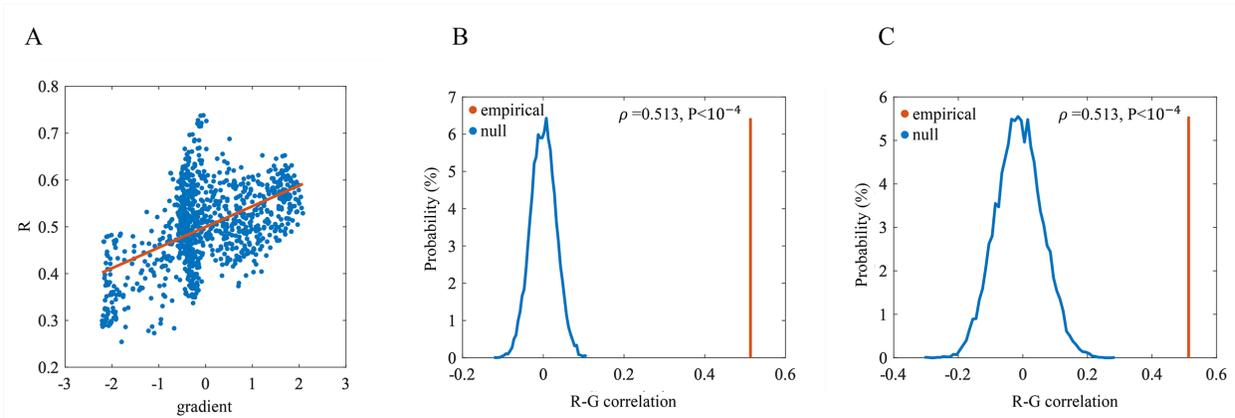

**Fig. 5. Convergent structure-function relationships along the unimodal-transmodal gradient.** (**A**), Coupling strength R estimated by high-frequency eigenmodes is positively correlated with the functional gradient. (**B**), The empirical correlation coefficient (red line) is significantly larger than null distribution generated by randomly permuting nodes' gradient positions (blue). (**C**), The empirical correlation coefficient (red line) is significantly larger than those obtained from artificial structural networks (blue).

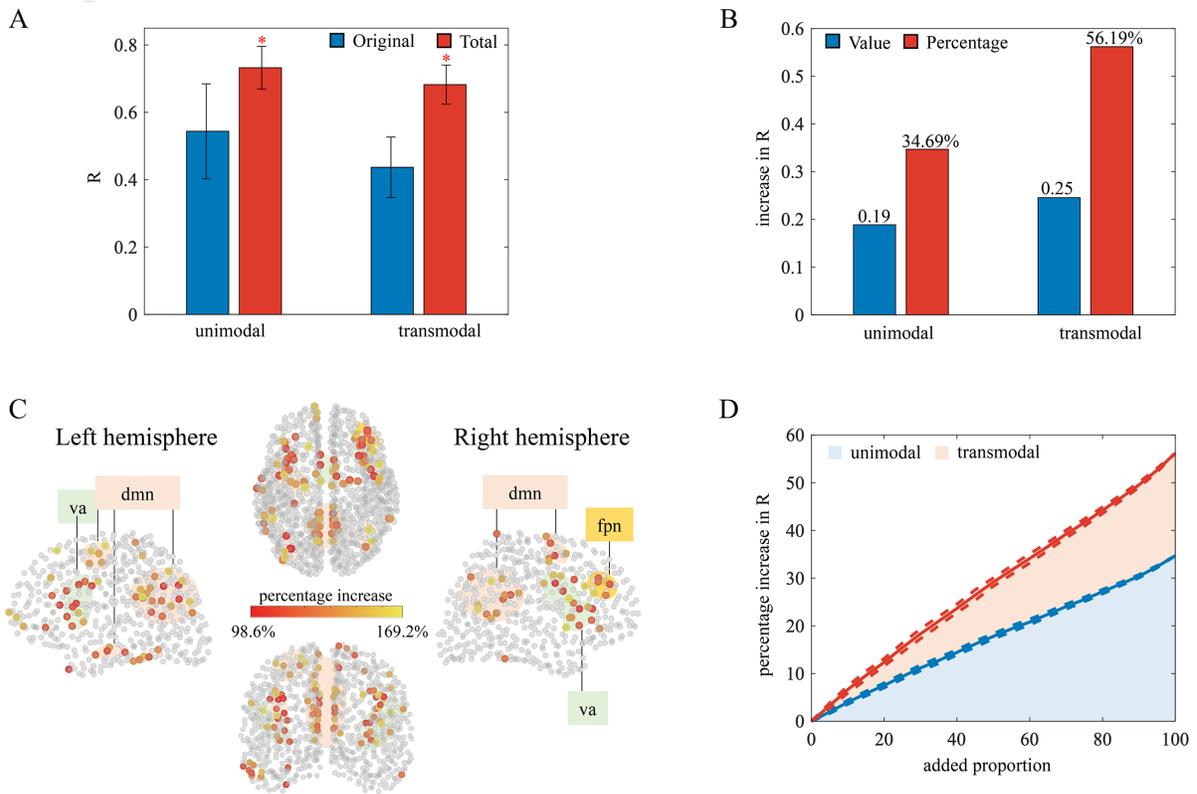

**Fig. 6. Enhanced structure-function tethering by high-frequency eigenmodes.** (**A**), Blue: Coupling strength between structure and function estimated by low-frequency predictors. Bed: Coupling strength estimated by both low-frequency and high-frequency predictors. Standard deviations are represented by error bars and statistically significant differences are marked by red stars (P < $10^{-3}$). (**B**), The average increments in prediction accuracy R for unimodal and transmodal groups are expressed as absolute (red) and relative (blue) values. (**C**), The spatial distribution of the top 10% of nodes with the highest percentage increases in R. (**D**), Growth curves of prediction accuracy R for unimodal and transmodal groups. The horizontal axis indicates the proportion of high-frequency eigenmodes added to the prediction model while the vertical axis is the percentage increase in R. At each adding step, the high-frequency eigenmode is randomly selected. The average percentage increase in R is shown in solid lines while standard deviations are expressed as dash lines (100 repetitions).

**Supplementary Materials**

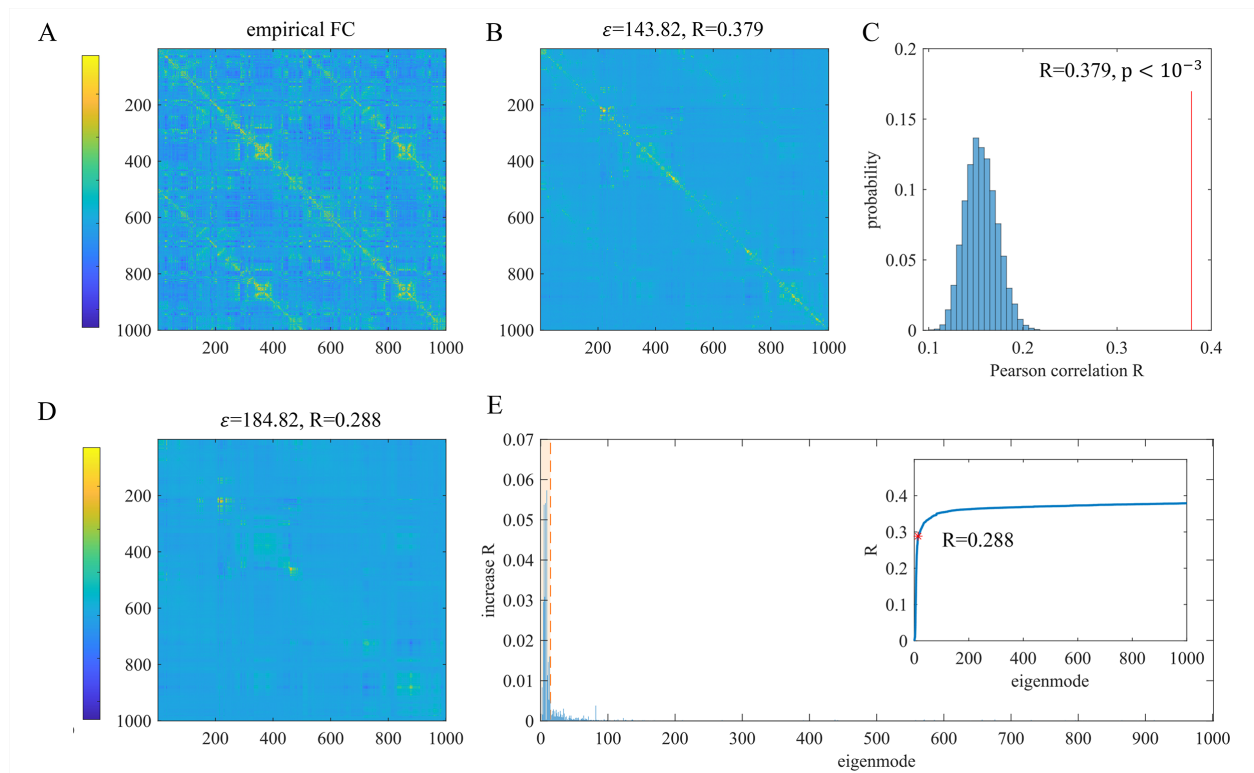

**Fig. S1.**

Extracting low-frequency eigenmodes. (A), The empirical functional connectivity matrix across 1,000 nodes. (B), The estimated functional connectivity matrix based on whole-brain regression model using all eigenmodes as predictors. The prediction performance is estimated as fitting errors $\varepsilon_{eigen}$ and Pearson correlation R between predicted and empirical functional connectivity. (C), Blue: the null distribution of prediction accuracy R generated by pseudo-eigenmodes. Red: the prediction accuracy R of empirical eigenmodes. (D), The estimated functional connectivity matrix predicted by low-frequency eigenmodes. (E), The histogram shows the contribution of individual eigenmodes to structure-function prediction. Orange area extracts low-frequency eigenmodes that have significant prediction contribution with respect to pseudo-eigenmodes. The inset figure shows the growth curve of prediction accuracy R with the number of eigenmodes. The red star indicates the prediction accuracy of low-frequency eigenmodes.

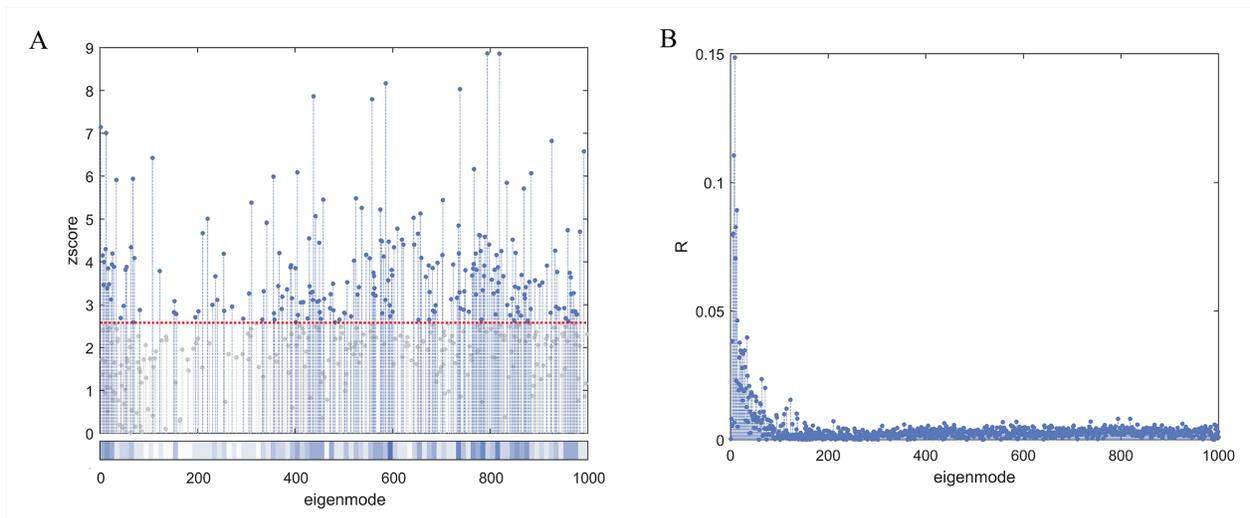

**Fig. S2.**
Prediction performance of single eigenmodes. (A), The significance of prediction contribution estimated by z scores transformed from R values. (B), The whole-brain prediction accuracy R estimated by each individual eigenmode.

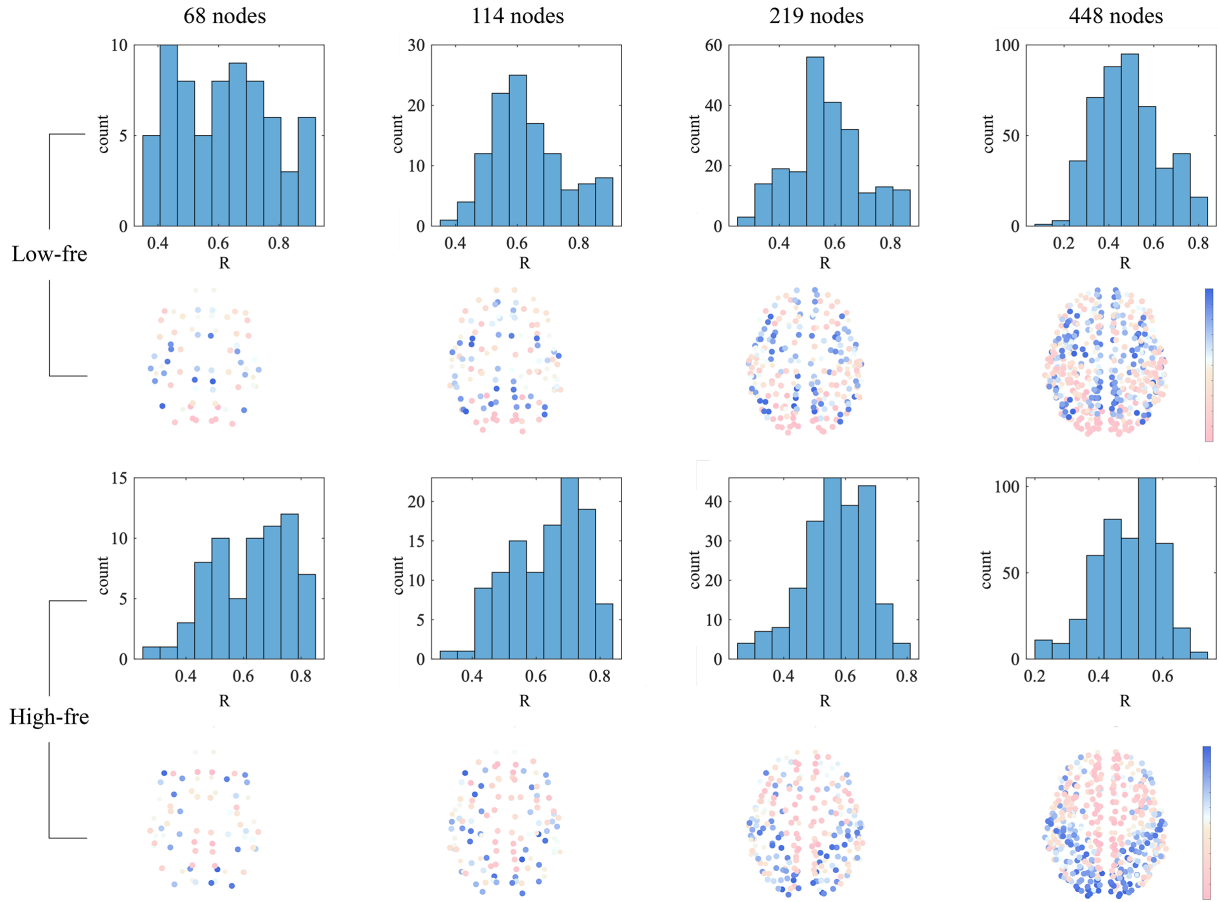

**Fig. S3.**

Consistency across spatial resolutions. The node-wise structure-function predictions based on low-frequency and high-frequency eigenmodes are respectively repeated in another four spatial resolutions (68, 114, 219, 448 nodes). The spatial patterns of structure-function R are visually similar.

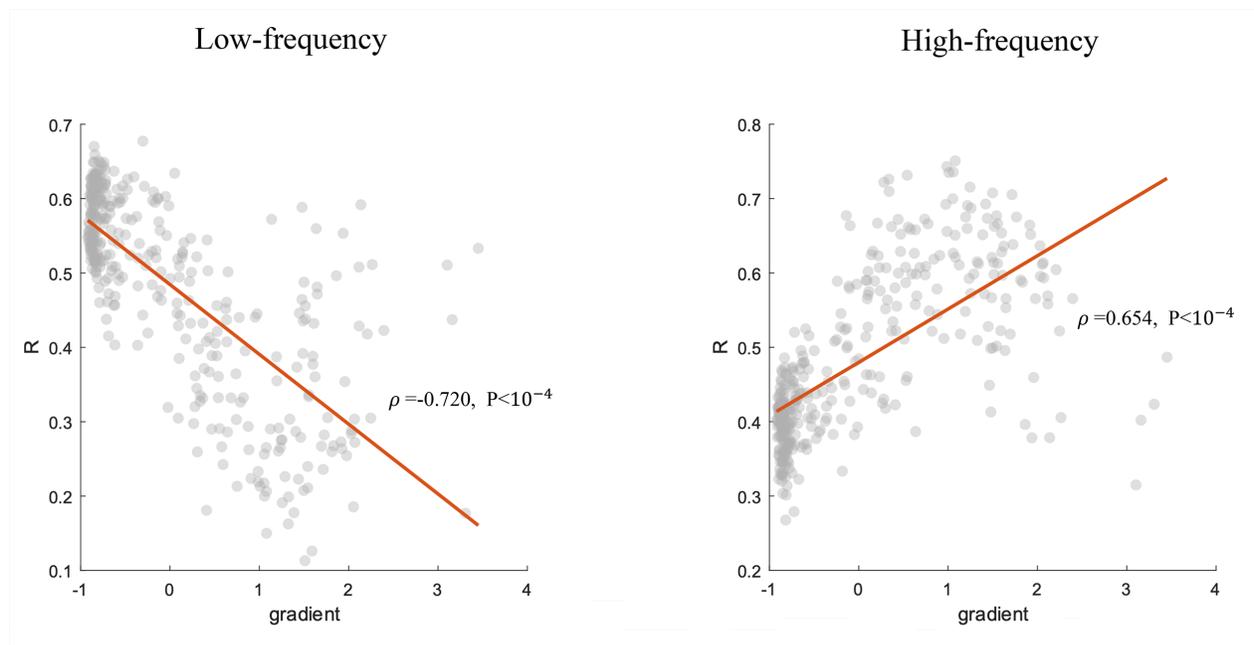

**Fig. S4.**
Verification on an independent dataset. The main results (structure-function divergence and convergence along the unimodal-transmodal gradient) are replicated in an independently collected dataset (Human Connectome Project HCP).

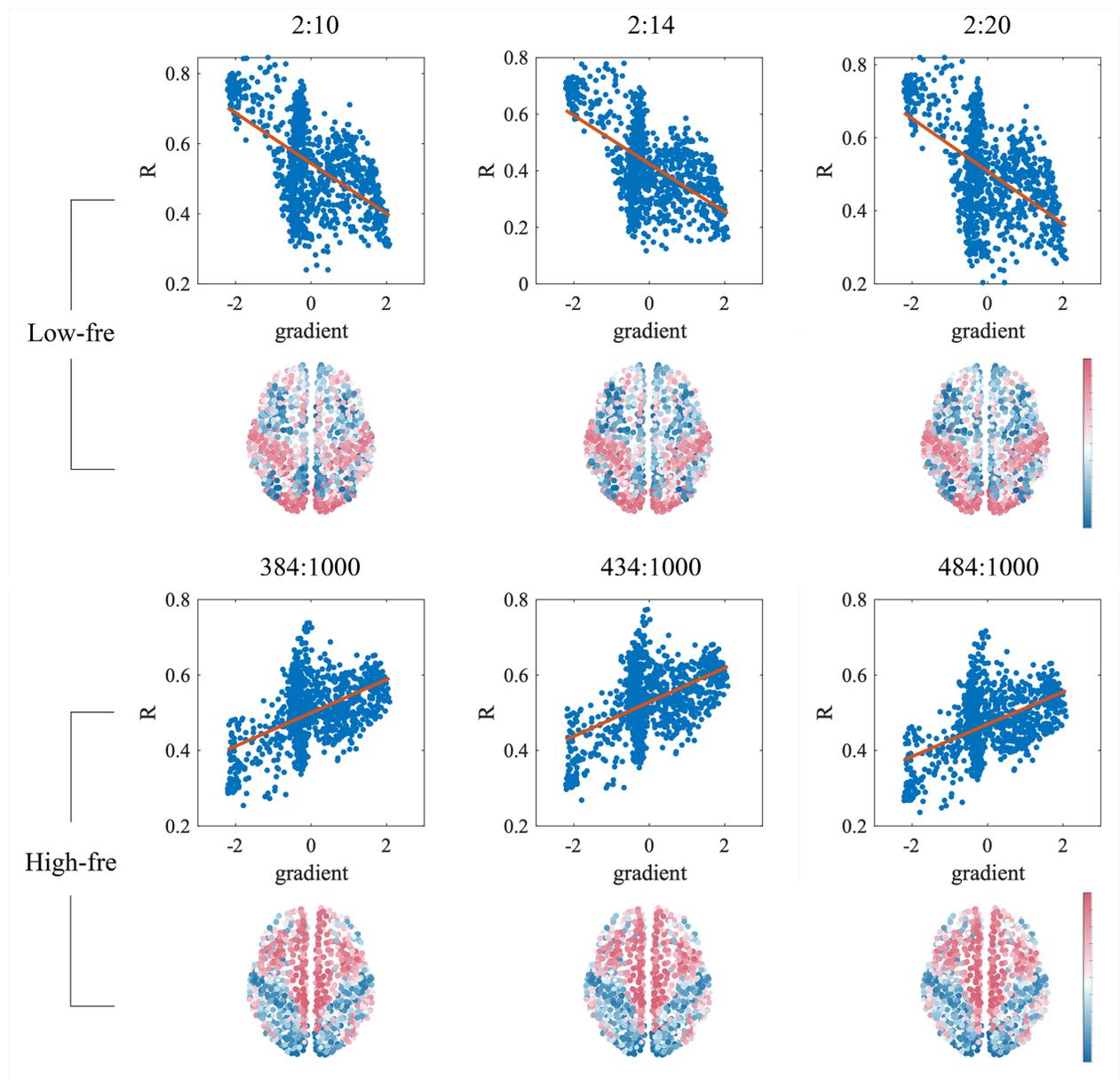

**Fig. S5.**

Robustness to frequency thresholds. The relationship between coupling strength R and the functional gradient is stable under different definitions of low-frequency and high-frequency eigenmodes.

Page **29** of **29**